\documentclass[aps,prc,twocolumn,showpacs,preprintnumbers,amsmath,amssymb,floatfix,nofootinbib]{revtex4-1}
\usepackage{graphicx}
\usepackage{bm}
\usepackage{pstricks}
\usepackage{color}
\usepackage{dcolumn}
\usepackage{amsmath}
\usepackage{bbm}

\definecolor{prel}{rgb}{1,1,1}
\begin{document}

\title{Appearance of the Single Gyroid Network Phase in Nuclear Pasta Matter}

\author{B. Schuetrumpf$^{1,2}$, M. A. Klatt$^3$, K. Iida$^4$, G. E. Schr\"oder-Turk$^3$, J. A. Maruhn$^1$, K.
Mecke$^3$, P.-G. Reinhard$^3$}

\affiliation{
$^1$Institut f\"ur Theoretische Physik, 
Goethe Universit\"at Frankfurt, D-60438 Frankfurt, Germany 
} 
\affiliation{
$^2$FRIB/NSCL Michigan State University, East Lansing, MI 48824, USA
} 
\affiliation{
$^3$Institut f\"ur Theoretische Physik, Friedrich-Alexander-Universit\"at Erlangen-N\"urnberg,
D-91058 Erlangen, Germany
} 
\affiliation{
$^4$Department of Natural Science, Kochi University, 2-5-1 Akebono-cho, Kochi
780-8520, Japan
} 

\date{\today}

\begin{abstract}
Nuclear matter under the conditions of a supernova explosion unfolds
into a rich variety of spatially structured phases, called nuclear
pasta. We investigate the role of periodic network-like structures with
negatively curved interfaces in nuclear pasta structures, by static
and dynamic Hartree-Fock simulations in periodic
lattices. As the most prominent result, we
identify for the first time the {\it single gyroid} network
structure of cubic chiral $I4_123$ symmetry, a well known configuration in
nanostructured soft-matter systems, both as a dynamical state and as a
cooled static solution.  Single gyroid structures form
spontaneously in the course of the dynamical simulations. Most of
them are isomeric states. The very small energy differences to the
ground state indicate its relevance for structures in nuclear pasta.
\end{abstract}


\maketitle

\section{Introduction}

Nuclear matter, although not observable in laboratories on earth,
plays a crucial role in astrophysical scenarios such as neutron stars
or core-collapse supernovae \cite{Bethe,Suzuki}. Near equilibrium
density, nuclear matter is a homogeneous quantum liquid, somewhat
trivial from a structure perspective.  However, an exciting world of
various geometrical profiles develops at lower densities covering
ensembles of rods, slabs, tubes, or bubbles
\cite{Ravenhall,Hashimoto,Pais,Watanabe2001,Schuetrumpf2013a}. Most
of these phases can be considered as manifestations of liquid crystals
\cite{Pethick1998} and the geometrical analogy to spaghetti, lasagna
etc. has led to summarize these under the notion of a nuclear
``pasta''. Their complex shapes and topologies can be classified by integral curvature measures
\cite{Nakazato2009,Nakazato2011,Schuetrumpf2013a}, developed in
the realm of soft matter physics and known as Minkowski
functionals~\cite{Mecke:1998,MeckeStoyan:2000,SchroederTurketal:2010,SchroederTurk:2013}.
  
\begin{figure}[tb]
  \centering
  \vspace*{0.55cm}
  \includegraphics[width=\linewidth]{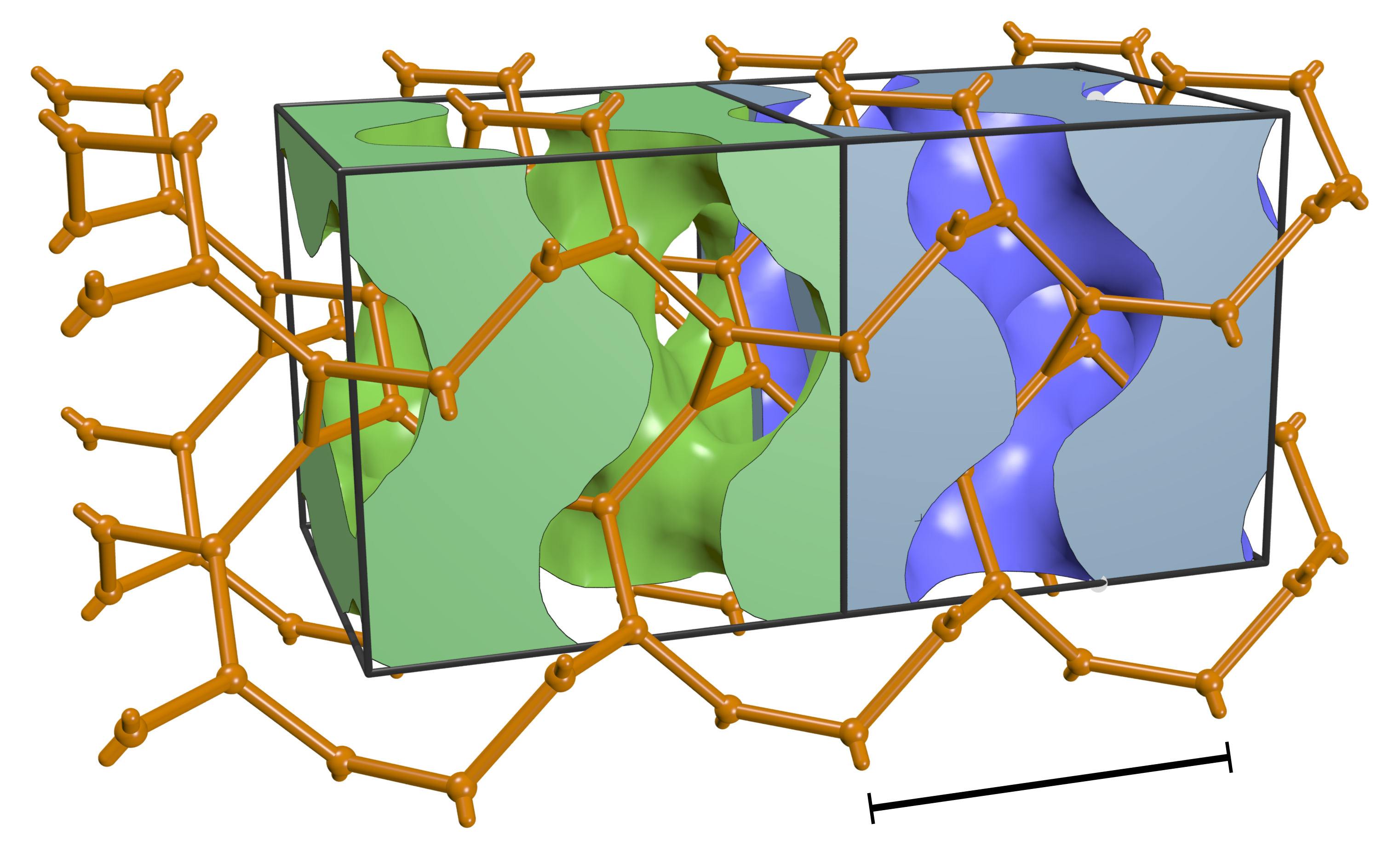}%
  \rput(-5.8,5.5){\large Pasta Matter}%
  \rput(-1.9,5.46){\large Single Gyroid}%
  \rput(-2.02,0.2){$22\,\mathrm{fm}$}%
  \caption{(Color online) Gyroidal pasta shape: the green structure on the left hand represents
the density distribution of the gyroidal state of nuclear pasta matter computed
with TDHF for an average density of $0.06{\rm\,fm^{-3}}$ and box length $a=22\,\mathrm{fm}$.
Shown is the Gibbs dividing surface with a corresponding threshold density. The solid volume representing
densities above this value and the void representing densities below this value. The blue structure on the right hand side shows the
nodal approximation \eqref{eq:gyroid} of a single gyroid CMC surface at the same volume
fraction. 
Also shown by orange bars is a gyroid network in the void phase of both the
pasta shape and the nodal approximation, showing that they are indeed
homotopic. 
Black frames are guides to the eye, of size $1.25\,a$ the cubic lattice parameter.}
    \label{fig:basic}
\end{figure}

A particularly intricate structure amongst the pasta phases is the
gyroid, a triply-periodic geometry consisting of two inter-grown
network domains separated by a periodic manifold-like surface which is
(at least on average) saddle-shaped and with negative Gaussian
curvature (cf. Fig.~\ref{fig:basic}). In soft-matter systems, these
periodic saddle-shaped surfaces have been found in solid biological
systems \cite{MichielsenStavenga:2008,
  SaranathanOsujiMochrieNohNarayananSandyDufresnePrum:2010,
  SchroederTurkWickhamAverdunkBrinkFitzGeraldPoladianLargeHyde:2011,
  GalushaRicheyGardnerChaBartler:2008,Pouya:11,Wilts2012,Nissen:1969},
in the so-called 'core-shell' gyroid phase of di-block copolymers
\cite{HajdukHarperGrunerHonekerKimThomasFetters:1994} and in inverse
bicontinuous phases in lipid-water systems
\cite{Larsson:1989}. Gyroid-like geometries can also be expected in nuclear pasta,
due to a balance between the nuclear and Coulomb forces
\cite{Nakazato2009,Nakazato2011}. Similar kinds of periodic
bicontinuous structures are discussed in supernova cores and neutron
star crusts \cite{Pethick,Mat06a}. It is generally accepted that
liquid crystalline phases, i.e., pasta phases, occur in supernova
cores in the form of slabs, rods and tubes
\cite{Ravenhall,Hashimoto,Pais,Watanabe2001}.  The search for
elaborate structures in astro-physical matter with self-consistent
nuclear models has a long history, starting from the first full
Skyrme-Hartree-Fock (SHF) simulation of \cite{Bonche}. With
continued refinement of the calculations, more and more intricate
structures had been discovered. For example, the possible occurrence
of periodic bicontinuous structures was found by stationary
Hartree-Fock calculations
\cite{Goe01aR,Mag02,Goe07a,NewtonStone,Pais}, later on in dynamical
simulations of supernova matter using time-dependent Hartree-Fock
(TDHF) calculations for supernova matter \cite{Sebille,Sebille2011}
and also in a quantum molecular dynamics approach
\cite{Sonoda2008}. Gyroids were examined so far only within a
liquid-drop model \cite{Nakazato2009,Nakazato2011} where double
gyroids were found to be energetically close to the ground state. (Note 
the important difference between single and double Gyroid geometries, see Fig.~\ref{fig:gyroids}.) If
realized in supernova matter, the network-like percolating nature of
the gyroid could greatly affect neutrino transport during the collapse
of a massive star's core and the subsequent core bounce. It is the aim
of this paper to investigate gyroid structures on the basis of fully
quantum-mechanical Hartree-Fock and TDHF simulations. To that end, we
employ the well established Skyrme-Hartree-Fock (SHF) energy
functional which provides a reliable description of nuclei and nuclear
dynamics over the entire nuclear landscape \cite{Bender03} and also in
astro-physical systems \cite{Stone2007}.

Figure~\ref{fig:basic} provides a graphical demonstration of the key
result of this article, namely the occurrence of a meta-stable gyroid
phase in nuclear pasta. It shows the Gibbs diving surface (see
section \ref{eq:cmc}) and draws the liquid phase as filled, the gas
phase as void (although in practice filled with some neutron dust).
The network-like domain on the left-hand side of the figure represents
the liquid (or high density) domain in full SHF calculations whose
shape and topology match closely those of one of the gyroid network
domains (shown on the left hand side of the figure). The remaining
void space (the gas phase) forms a complementary network-like domain
with the same topology, albeit of different volume fraction.

\section{Constant mean-curvature (CMC) surfaces}
\label{eq:cmc}

A quantitative description and identification of the domain shapes in
nuclear pasta matter is afforded by the so-called Minkowski
functionals, that are here evaluated for the isodensity surfaces of
the nuclear density field $\rho(\mathbf{r})$.  We first deduce
a surface from the given density which divides the space into a liquid
phase (solid), and a gas phase (void).  To that end, we use the Gibbs
dividing surface (a specific isodensity surface), a standard tool of
solid state physics \cite{Shchukin}, which is chosen such that the
liquid phase volume $V_b$ contains all the matter in a liquid phase
with constant density $\rho_0$. 
This means that
$V_b\rho_0=N_\mathrm{tot}$ where $N_\mathrm{tot}$ is the total
number of particles in the box. For $\rho_0$ we take the maximum
density of the individual states. The fraction $u$ of liquid volume is

\begin{equation}
  u=\frac{V_b}{V}=\frac{\rho}{\rho_0}
  \;,\; 
  \rho=\frac{N_\mathrm{tot}}{V}
\end{equation}
where $\rho$ is the mean density and $V$ the volume of the whole
numerical box.

Having defined a surface, we can compute the Minkowski functionals
which in three dimensions are the volume $V_b$, the surface area
$A$, the integrated mean curvature $\int H{\rm\,d}A$, and the Euler
characteristic $\chi=1/(4\pi)\int{K}{\rm\,d}A$ \cite{Schuetrumpf2013a}, a
topological constant only taking integer values \cite{bibfootnote}. Mean curvature
and $\chi$ are computed the following way: each point on a surface
in 3-space has two principal curvatures $\kappa_1$ and $\kappa_2$;
these are used to compose the mean curvature as
$H=(\kappa_1+\kappa_2)/2$ and the Gaussian curvature as
$K=\kappa_1\kappa_2$. Negative values of $\chi$
indicate
network-like structures \cite{Evans2013}. We are interested in what is
called {\em hyperbolic} surfaces, where the interface is saddle-shaped
and has Gaussian curvature $K\le 0$ ($\kappa_1$ and $\kappa_2$ with
different sign). Such interfaces can form the continuous bounding
surfaces of {\em periodic} labyrinth-like domains; they now have a
firm place in the taxonomy of soft-matter nanostructures
\cite{HydeLanguageOfShape:1997}. More specifically, we search within
the class of constant-mean-curvature surfaces (CMC) which have
constant $H$.  They provide structures with the same topology and
symmetries, yet with variable volume fractions $u(H)$
\cite{GrosseBrauckmann1997,GrosseBrauckmann1997b,AndersonEtAl1987}.
Gyroids belong to the CMC and require, in particular, that $\chi=-4$.

\begin{figure}[tb]
  \centering
  \includegraphics[width=\columnwidth]{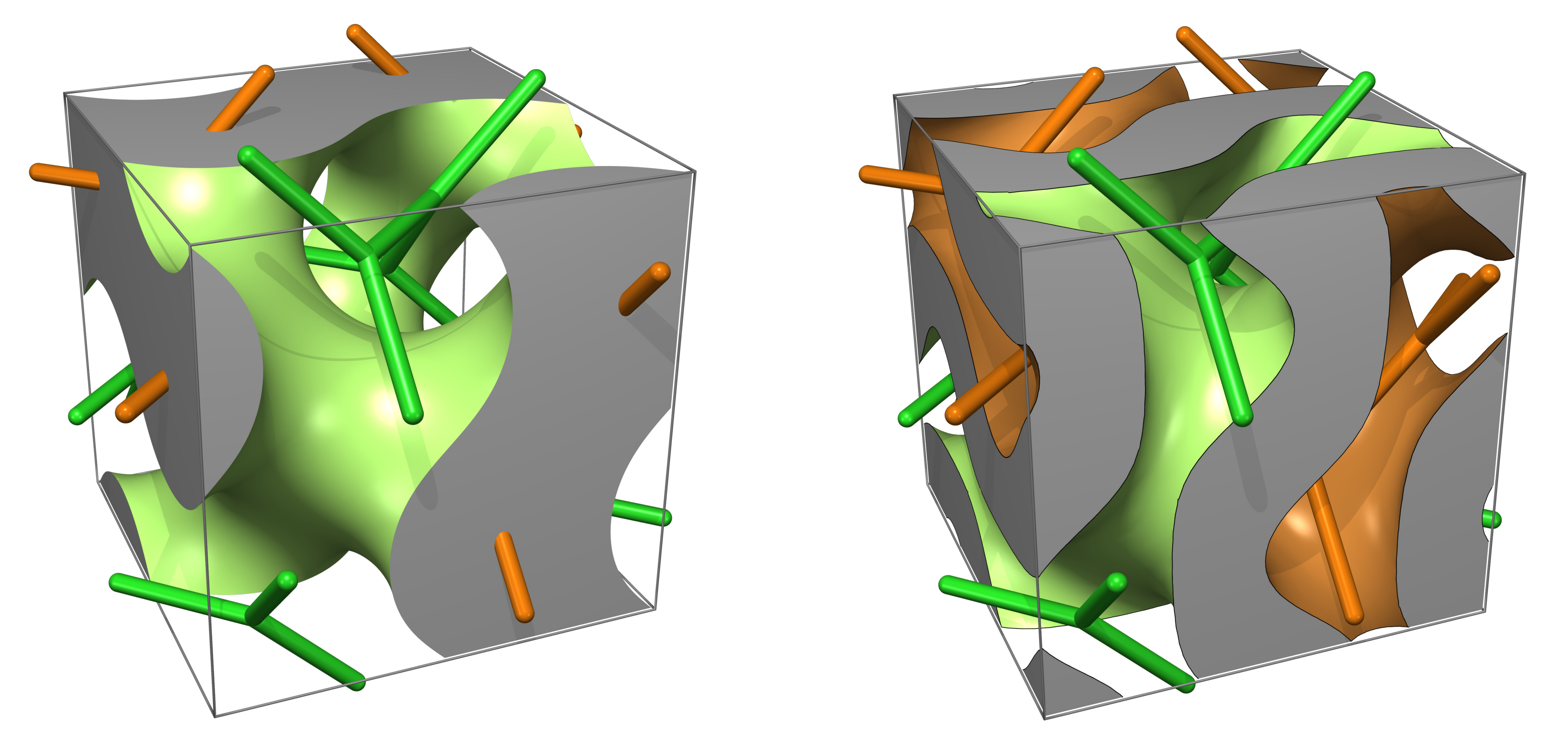}
  \caption{(Color online) Distinction between the {\it single gyroid} (left) and
{\it double gyroid} (right) geometries. The single gyroid consists of a single
solid domain (gray) and a single void domain, separated by CMC Gyroid surface.
The double gyroid consists in a solid domain (gray) represented by a sheet of
finite width draped onto the Gyroid minimal surface and two distinct void
domains, each forming a network-like labyrinth. The interfaces between the solid
and the void domain in the double Gyroid are two distinct surfaces (brown and
green), modelled e.g. by CMC surfaces with constant curvatures $\pm h_0$. A
second form of the double gyroid, where solid and void is interchanged, also
exists. (images from Ref.~\cite{Kapfer:2011-biomat})}
  \label{fig:gyroids}
\end{figure}

In soft-matter systems, these periodic saddle-shaped surfaces occur in two
forms, called ``single'' or ``double'', see Fig.\ref{fig:gyroids}. The {\em double gyroid} (DG) is a
structure composed of two inter-grown non-overlapping network domains, bounded by
two CMC gyroid surfaces with mean curvatures $\pm H$ separated by the so-called
matrix phase. The {\em single gyroid} (SG, or simply G) is composed of two
domains, one solid and one void of volume fractions $u$ and $(1-u)$, with a
gyroid CMC surface as the interface between them; SG have
been identified so far only in solid biological systems
\cite{MichielsenStavenga:2008,
SaranathanOsujiMochrieNohNarayananSandyDufresnePrum:2010,
SchroederTurkWickhamAverdunkBrinkFitzGeraldPoladianLargeHyde:2011,
GalushaRicheyGardnerChaBartler:2008,Pouya:11,Wilts2012,Nissen:1969}. We
concentrate in the following on the ``single'' structures, because double
structures were not found to be stable for the assumed parameters in this work.

However, G is not the only CMC structure meeting all of the above
criteria. They come along with the related primitive (P) and diamond (D)
structures which all three are distinguished by their structural uniformity or
homogeneity \cite{Hyde:1990}. While any negatively curved interface necessarily
exhibits spatial variations of the point-wise Gaussian curvature $K$ (see Ref.~\cite{Evans:2014} for a discussion) and the
point-wise domain width or thickness $d$, the P, D and G surfaces appear to be
optimal structures with minimal variations of $K$ \cite{FogdenHyde:1999}, and
the G surface is the optimal structure with minimal variations of domain width
$d$ \cite{SchroederFogdenHyde:2006,SchroederRamsdenChristyHyde:2003}. This
homogeneity is likely to explain the ubiquity of these three structures in
soft-matter systems, particularly that of the G.
The close relationship between the P, D and G is also visible from
the fact that they can be connected by the Bonnet
transformation~\cite{HydeLanguageOfShape:1997} which leaves all metric and
curvature properties unchanged and which relates the corresponding unit cell box
lengths as $a_P/a_G=0.81$ and $a_D/a_G=1.27$. (These ratios correspond
to the oriented space groups for the single structures with symmetry groups
$Pm\bar{3}m$, $Fd\bar{3}m$, and $I4_132$ for the P, D, and G structures.) While
it is not a physical transformation (as intermediate structures are
self-intersecting and not embedded), the specific 'Bonnet ratio' of lattice
parameters as given above is often observed in lipid systems that form two of
the P, D and G mesophases \cite{BaroisHydeNinhamDowling:1990}. This Bonnet ratio
can also help to relate and characterize structures. The D surfaces
are not found to be stable and thus ignored here.

\begin{table*}[t]%
  \caption{\label{tab:volfrac_minkfunc}
Surface area $A$ and mean curvature $H$
of the exact CMC and pasta P and G surfaces for different mean densities $\rho$
and corresponding volume fractions $u$. For the pasta shapes the average mean
curvature $\int H{\rm\,d}A/A$ is given. The values $t_P$ and $t_G$ are the
threshold values for the nodal representations of P and G in
Eqs. (\ref{eq:primitive}) and (\ref{eq:gyroid}) with volume fractions $u$. The
unit of length is the lattice constant.}
  \centering
  \begin{ruledtabular}
  \begin{tabular}{D{.}{.}{-1} d @{\hspace{0.5cm}} d d @{\hspace{0.5cm}} d d@{\hspace{0.5cm}} d @{\hspace{1cm}} d d @{\hspace{0.5cm}} d d @{\hspace{0.5cm}}d}
     &  & \multicolumn{5}{c@{\hspace{1cm}}}{Primitive}  & \multicolumn{5}{c}{Gyroid} \\
    \multicolumn{1}{c}{$\rho[{\rm\,fm^{-3}}]$} & \multicolumn{1}{c@{\hspace{0.5cm}}}{$u$} & \multicolumn{2}{c@{\hspace{0.5cm}}}{$A_{P}$} & \multicolumn{2}{c@{\hspace{0.5cm}}}{$H_{P}$} & \multicolumn{1}{c@{\hspace{1cm}}}{$t_P$} & \multicolumn{2}{c@{\hspace{0.5cm}}}{$A_{G}$} & \multicolumn{2}{c@{\hspace{0.5cm}}}{$H_{G}$} &\multicolumn{1}{c}{$t_G$} \\
     & & \multicolumn{1}{c}{CMC} & \multicolumn{1}{c@{\hspace{0.5cm}}}{pasta} & \multicolumn{1}{c}{CMC} & \multicolumn{1}{c@{\hspace{0.5cm}}}{pasta} & & \multicolumn{1}{c}{CMC} & \multicolumn{1}{c@{\hspace{0.5cm}}}{pasta} & \multicolumn{1}{c}{CMC} & \multicolumn{1}{c@{\hspace{0.5cm}}}{pasta}&\\
    \hline
  0.0326 & 0.22 &      &      &       &       & 0.99  & 2.61 & 2.84      & 1.96  & 1.85  & 0.86     \\
  0.0434 & 0.29 & 2.11 &      & 1.22  &       & 0.75  & 2.83 & 3.01      & 1.31  & 1.31  & 0.66     \\
  0.0543 & 0.37 & 2.26 & 2.46 & 0.66  & 0.60  & 0.46  & 3.00 & 3.25      & 0.74  & 0.75  & 0.41     \\
  0.0651 & 0.45 & 2.33 & 2.51 & 0.24  & 0.23  & 0.18  & 3.08 & 3.34      & 0.28  & 0.26  & 0.16     \\
  0.0715 & 0.50 & 2.35 &      & 0.0   &       & 0.0   & 3.09 &           & 0.0   &       & 0.0      \\
  0.0759 & 0.53 & 2.34 & 2.54 & -0.13 & -0.17 & -0.09 & 3.09 & 3.35      & -0.15 & -0.24 & -0.08    \\
  0.0868 & 0.61 & 2.29 & 2.44 & -0.55 & -0.55 & -0.39 & 3.02 & 3.30      & -0.62 & -0.72 & -0.34    \\
  0.0977 & 0.70 & 2.15 &      & -1.08 &       & -0.69 & 2.87 & 3.10      & -1.19 & -1.32 & -0.60  
  \end{tabular}
  \end{ruledtabular}
\end{table*}

Tab.~\ref{tab:volfrac_minkfunc} shows values for surface area $A$, volume
fraction $u$ and mean curvature $H$ for the G and P surface. One can represent
the various surfaces by the simple {\em nodal} approximation
\cite{NesperSchnering1991} in terms of ``potential'' functions $\phi_i(x,y,z)$
with $i\in\{ {\rm S,P,G}\}$.  For structures scaled to lattice parameters
$a=2\pi$ one uses
\begin{eqnarray}
  \phi_S/\phi_0&=&\cos x \quad,
\label{eq:slab}\\
  \phi_P/\phi_0&=&\cos x + \cos y +\cos z \quad,
\label{eq:primitive}\\
  \phi_G/\phi_0&=&\cos x \sin y + \cos y \sin z + \cos z \sin x\quad,
 \label{eq:gyroid}
\end{eqnarray}
where $\phi_0$ is some constant potential value.  The slab (S) is
generated by the potential (\ref{eq:slab}) producing parallel sheets
of matter, also called ``lasagna''.  The P surface from potential
(\ref{eq:primitive}) consists in a simple cubic lattice of rods in
three orthogonal directions which have common crossing points.  The G
surface is produced by Eq.  (\ref{eq:gyroid}). The minimal surfaces
are approximated by $\phi_i(x,y,z)=0$ and the CMC surfaces by
$\phi_i(x,y,z)/\phi_0=\pm t$ (where $t$ is the threshold value
  for the nodal representation, see table
  \ref{tab:volfrac_minkfunc}). For the volume fractions considered
here, the differences between the nodal approximations and the exact
CMC surfaces are negligible compared to the spatial resolution of the
computational grid.

\section{Summary of numerical simulation scheme}

Here we use nuclear (TD)HF calculations
employing the Skyrme energy-density functional, for a review see
Ref.~\cite{Bender03}. To maintain consistency with our previous
study \cite{Schuetrumpf2013a}, we use the parametrization SLy6
\cite{Chabanat} and consider the proton fraction $X_p=1/3$ throughout.
The calculations are performed on an equidistant grid in 3D coordinate
space. Box size and lattice spacing are adjusted to give the desired 
conditions of matter. We consider grid spacings of $0.875 {\rm\,fm}
\leq\Delta r\leq 1.125 {\rm\,fm}$ with $N_x=N_y=N_z$ grid points
ranging from 16 to 24. This spans periodic simulation boxes in
  the range $a=15-26{\rm\,fm}$.

Static HF solutions are computed by accelerated gradient iteration
\cite{Rei82n} while time stepping is done with Taylor expansion of the
mean-field time-evolution operator, for technical details see
\cite{Mar14a}.  We use the Coulomb solver with periodic boundary
conditions. Charge neutrality is enforced by assuming a
compensating homogeneous cloud of negative charges. This
approximation ignores electron screening. Its effect on matter under
astro-physical conditions have been much discussed in the past
\cite{Mar05a,Watanabe2003a,Dorso2012,Alcain}.  Although screening
lengths vary widely and reach occasionally the order of structure
size, the net effect of electron screening is found to be small
\cite{Mar05a,Watanabe2003a}, thus excusing the homogeneous
approximation for the present exploration.

The search for locally stable CMC configurations is done by
biased initialization. First, we initialize the system with plane
waves up to the wanted amount of protons and neutrons.  For the
first 1000 static iterations, we imprint a bias by adding an
external guiding potential $\phi_i$ according to
Eqs.~(\ref{eq:slab})-(\ref{eq:gyroid}) with $\phi_0=10
{\rm\,MeV}$. After these first 1000 iterations, we switch off the
guiding potential and continue with 9000 further pure HF
iterations. The system is then driving to a local minimum, not
necessarily the ground state. If the stationary state thus found
stays close to the intended structure, we consider this structure as
(meta)stable.

The dynamical TDHF simulations are initialized similar as in
Ref.~\cite{Schuetrumpf2013a}.  We take ground-state wavefunctions of
$\alpha$-particles and place them randomly in space. As we are
simulating at proton fraction $X_p=1/3$ only half of the neutrons
can packed in these $\alpha$-particles.  The remaining neutrons are
added as plane waves filling the states with lowest kinetic
energy. After all, the set of wavefunctions is ortho-normalized. In
contrast to \cite{Schuetrumpf2013a}, the $\alpha$-particles were taken
at rest and the minimal distance between them was larger such that the
distribution is more homogeneous (yet still random) and less
energetic. Nonetheless, this initialization scheme produces systems
at rather large excitation energies with temperature
$T\approx7{\rm\,MeV}$.  The mean density was fixed to
$\rho=0.06{\rm\,fm^{-3}}$.  The box length was varied here in the
range $a=20-24{\rm\,fm}$. For each $a$, ten simulations were performed
running over $1000\,{\rm fm/c}$.  The final structure emerges typically
after $400\,{\rm fm/c}$ and then stays stable, of course, as a fluctuating state.

A word is in order about the analysis of sub-structures of
infinite systems in a finite simulation box. It is known that the
thus imposed periodic boundary conditions have an impact on the
structures due to possible symmetry violation by the box and
subsequent spatial mismatch \cite{Hoc81aB,All87,Oka12a}. A study within
classical molecular dynamics without Coulomb interactions implies
that this may be particularly critical for extended, inhomogeneous
structures \cite{Gimenez}.  Spurious shell effects may further blur
the analysis in a quantum-mechanical framwork \cite{NewtonStone}.
However, the numerical expense of fully quantum-dynamical
simulations sets limits on the affordable box sizes. To get some
idea on the size dependence, we vary here the box in the rather
broad range 16--26 fm.  This average smooths some artifacts and
thus allow even more rigorous structure identification. And yet, our
calculations are to be viewed as providing indicators of the
appearance of the single gyroid. It motivates us to plan much larger
calculations which, however, will run on a long time scale.

\section{Stationary structures}

First, we discuss results from static HF calculations under various conditions
of density and box size. The simulation of Ref.~\cite{Schuetrumpf2013a} used
only boxes with a box length of $a=16 {\rm\,fm}$. The most involved structure
found there is topologically identical to a P surface. Here we investigate
larger systems up to $a=26{\rm\,fm}$, all at the same proton fraction
$X_P={1}/{3}$.

\begin{figure}[tb]
  \centering
  \includegraphics[width=\columnwidth]{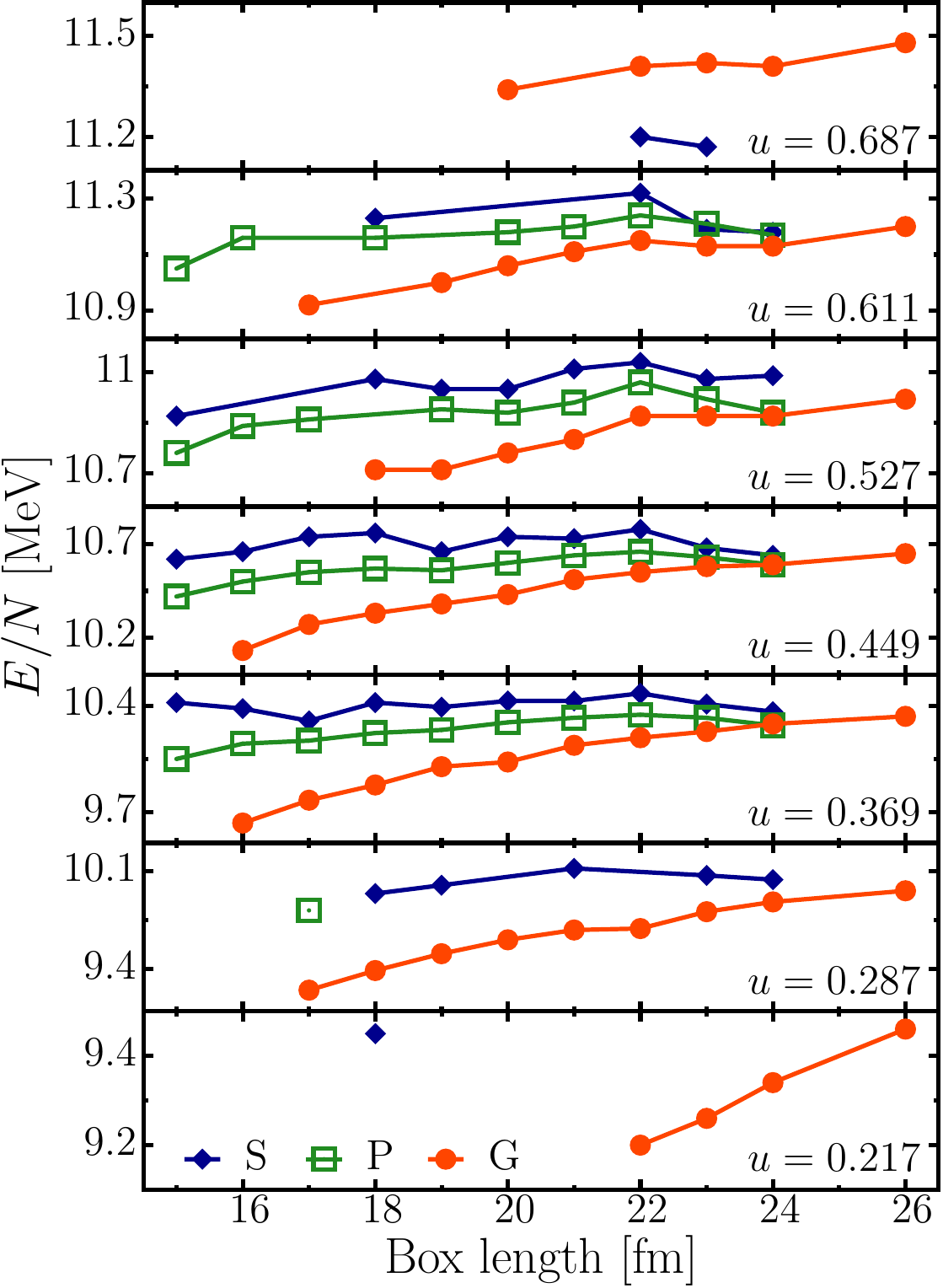}
  \caption{(Color online) Binding energies per nucleon $E/N$ for the metastable ground
states for different volume fractions. G (gyroid, dots), P (primitive, squares)
and S (slab, diamonds) denotes the initial guiding potential.
The states shown in this plot remained topologically stable for the 9000
iterations. Higher values of $E/N$ correspond to more tightly bound and 
hence more favorable solutions.}
  \label{fig:potcalcs}
\end{figure}

We found stable S, P and G structures, the latter both of ``single'' type. The
resulting energies are summarized in Fig.~\ref{fig:potcalcs}. The G is stable
for $a\geq22{\rm\,fm}$ in the widest range of volume fractions compared to P and
S. Binding energies of P and G show a clear scaling with box length. For P, the
binding energy has a maximum at $a\approx22{\rm\,fm}$. Under the assumption that
an effective description of the pasta Hamiltonian by curvature terms is
possible, we expect that the G has a maximum binding energy at
$a\approx27{\rm\,fm}$ due to the Bonnet transformation. The binding energies for
P at $a=22{\rm\,fm}$ and G at $a=26{\rm\,fm}$ are lower (and hence less
favorable) than for S, except for $u=0.687$, but at this high volume fraction
the rod(2) bubble shape (not shown here) has larger binding energies than both G
and S. Note that the DG, whose energy was calculated in the liquid drop model
\cite{Nakazato2009,Nakazato2011}, is unstable in the TDHF simulations.

Table~\ref{tab:volfrac_minkfunc} shows that surface area $A$ and mean curvature
$H$ of the pasta G and P surfaces agree with the exact values, within the rough
voxelization of the TDHF calculations. We note a slight but systematic deviation
of G and P obtained in TDHF from the nodal models: an anisotropic deformation of
the interface surface. While the nodal surface models have cubic symmetry, the
directional distribution of interfaces in TDHF are slightly biased towards an
axis, exposing more interfacial area in this direction that in the two
perpendicular ones. This can be quantified by Minkowski tensor shape analysis,
as detailed in \cite{SchroederTurketal:2010,SchroederTurk:2013}. The eigenvalue
ratio $\beta_{1}^{0,2}$ of the interface tensor $W_1^{0,2}$ (as defined in
\cite{SchroederTurk:2013}), evaluated for a Marching cubes representation of the
Gibbs Dividing surface interface,  adopts values not below 0.75 for G and values
between 0.4 and 1 for P structures.

\section{Dynamical stability}

In order to check dynamical formation and stability of the structures,
we performed TDHF simulations and use rather large excitation
energies ($T\approx7{\rm\,MeV}$) which provide a critical
counter-check.
\begin{figure}[tb]
\centerline{
\includegraphics[width=0.9\columnwidth]{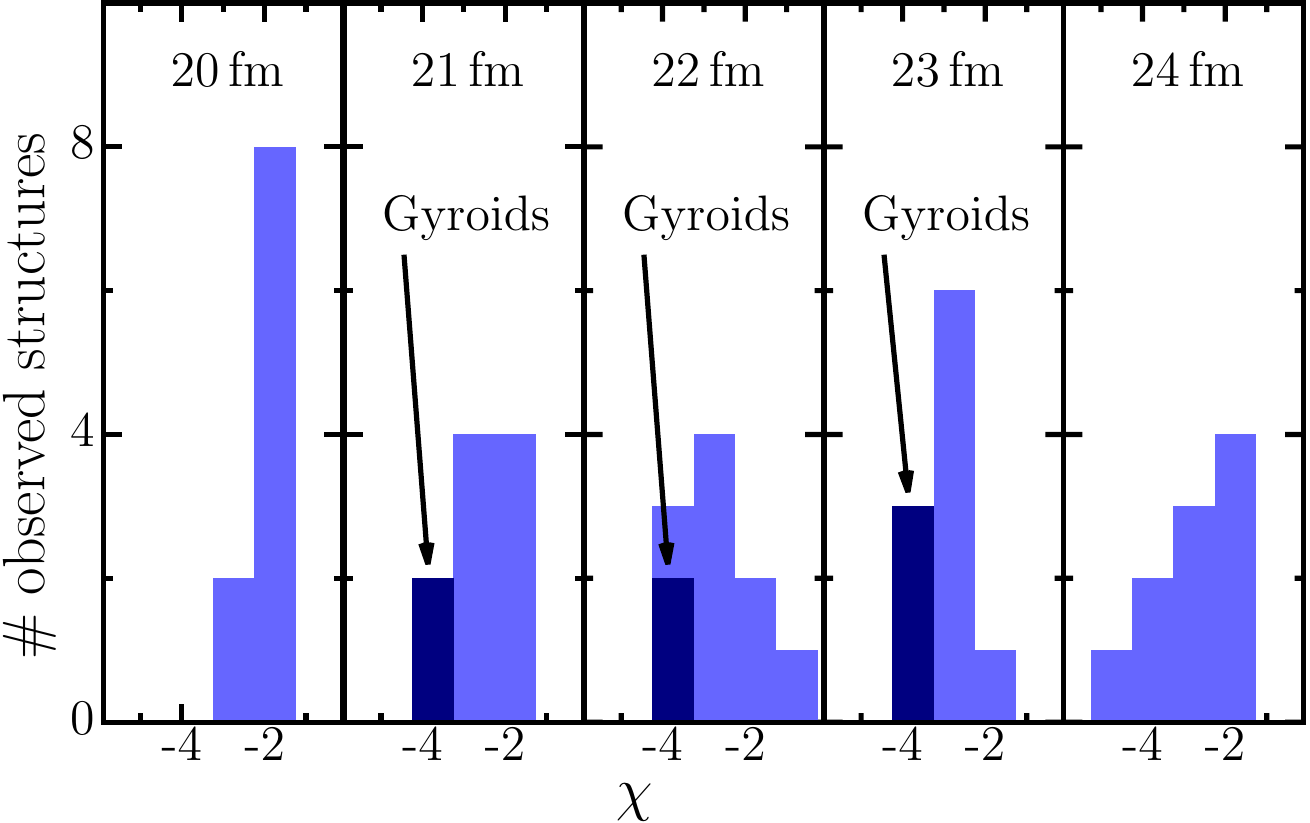}
}
\caption{(Color online) Histograms of the resulting Euler characteristic $\chi$
for different box lengths $20-24{\rm\,fm}$ of the dynamic, randomly initialized
calculations. The shaded area marks the identified gyroidal structures.}
\label{fig:histo}
\end{figure}
Figure~\ref{fig:histo} shows the histograms of the Euler
characteristic $\chi$ for all box lengths.  All structures have
negative Euler characteristic, implying that these are labyrinth-like
structures: $\chi=-2$ indicates P, $\chi=-3$ corresponds to mixed
structures not discussed here, $\chi=-4$ is a necessary condition for
G. We finally identify a G by trying to draw the G nodal network
through the void. If this turns out to be possible, we have
successfully found a single G structure. We have marked it
explicitly by shaded areas in figure~\ref{fig:histo}. The
figure demonstrates that a great variety of CMC structures can
emerge spontaneously in a finite temperature calculation. We note in
particular the repeated appearance of the involved, labyrinth-like
single gyroid (G) which thus shows that it can be stable even under
the extreme conditions of hight temperature. As quantum shell
effects disappear at temperatures $T>2$ MeV \cite{Bra81a}, we also
can conclude that the appearance of a G structure is not determined
by shell effects (may it be physical or spurious ones
\cite{NewtonStone}).  Note that 10 random samples per box do not
provide sufficient statistics to compare quantitatively the
abundances of structures. The figure merely indicates their possible
appearance even under these highly excited conditions.

The dynamical simulations thus have delivered a couple of (highly
excited) G structures. It is interesting to recover stationary G
starting static HF from the dynamical states. To that end, we take
final states of a TDHF simulation as initial states for a SHF ground
state iteration. Doing this for the dynamical G configurations,
three cases out of seven remain gyroidal.
The binding energies of the cooled gyroidal structure are close
to those from purely static calculations.  E.g., for $a=22{\rm\,fm}$
we find $E/N=10.357{\rm\,MeV}$ from cooling the dynamical
state versus $E/N=10.377{\rm\,MeV}$ from purely static HF.
However, the ``cooled'' dynamical G is more anisotropic (as seen
from the tensorial Minkowski measures)
while its scalar Minkowski converge to those of the statically
calculated G. The binding energy thus depends primarily on the scalar
measures like surface area, curvature, and bulk volume and is
rather insensitive to the deformation. We thus are probably dealing
with a variety of isomeric G configurations. This is another
interesting detail calling for further investigations.

\section{Conclusions}

We have investigated the appearance of non-homogeneous structures
in nuclear matter under astro-physical conditions, paying particular
attention to structures with network-like geometry and topology, 
amongst them as particularly appealing shape the single
Gyroid (G). To that end, we used static optimization as well as
dynamical simulations within a self-consistent nuclear mean-field
model (Skyrme-Hartree-Fock). The emerging structures have been
characterized by Minkowski measures. To uniquely identify a G, an
additional subsequent graphical analysis has been performed.  We have
shown that single gyroid (G) structures indeed emerge in static and
dynamical simulations. We have looked at further CMC surfaces and
find close competition with the primitive (P) surface and the slab
(S) while the diamond seem to play no role.  The static
calculations show that G and P are mostly metastable while the S
usually provides the ground state, however, with only slightly larger
binding energies. G and P, being triply periodic saddle surfaces, 
have maxima for the binding energies as function of
box length, for P at $a\approx22{\rm\,fm}$ and for G predicted at
$a\approx27{\rm\,fm}$, assuming that the Bonnet transformation is
applicable in this system. Dynamical simulations at high
excitation energy produce several cases where these structures
appear again. This indicates that these network-like geometries are rather robust
in nuclear matter.  We also find some static G structures with
deformations which may be due to the fact that the surface energy is
rather small under the given conditions thus easily allowing
deformed G isomers.

The large expense of these microscopic calculations limits presently
the size of the affordable numerical box. This inhibits so far a
unambiguous assessment by studying the trends with box size in
larger ranges. The present results are, however, strong indicators
for the appearance of CMC and particularly G structures which
call for further studies.

\begin{acknowledgments}
This work was supported by the Bundesministerium f\"ur Bildung und Forschung
(BMBF), contract 05P12RFFTG, the German science foundation (DFG) through the 
research unit ‘Geometry and Physics of Spatial Random Systems’, grant
ME1361/11, and by Grants-in-Aid for Scientific Research on Innovative Areas
through No. 24105008 provided by MEXT. The calculations have been performed on
the computer cluster of the Center for Scientific Computing of J. W.
Goethe-Universit\"at Frankfurt. B.S., K.I., and J.A.M. acknowledge the
hospitality of the Yukawa Institute for Theoretical Physics, where this work was
initiated. K.I. is grateful to K. Nakazato and K. Oyamatsu for useful
discussion.
\end{acknowledgments}
\bibliography{gyroid}
\end{document}